\documentclass[
12pt,
preprint,
onecolumn,
showpacs,
amsmath,
amssymb
]{revtex4}

\usepackage{graphicx}
\usepackage{dcolumn}
\usepackage{bm}

\newcommand{\hide}[1]{}
\newcommand{\eq}[1]{Eq.\,(\ref{#1})}

\newcommand{\noeq}[1]{(\ref{#1})}
\newcommand{\fig}[1]{Fig.~\ref{#1}}
\newcommand{\figs}[2]{Figs.~\ref{#1} and~\ref{#2}}

\newcommand{\tab}[1]{Table~\ref{#1}}

\newcommand{\viceversa}{{\em vice versa}}

\def\+#1{\ifmmode{{#1}^{\dagger}}\else{${#1}^{\dagger}$}\fi}
\def\bra#1{\ifmmode{\left< #1 \right|}\else{$\left< #1 \right|$}\fi}
\def\ket#1{\ifmmode{\left| #1 \right>}\else{$\left| #1 \right>$}\fi}
\def\braket#1#2{\ifmmode{\left< #1 | #2 \right>}\else{$\left< #1 | #2 \right>$}\fi}
\def\half{\ifmmode{\frac 1 2}\else{${1\over2}$}\fi}
\newcommand{\lsimeq}{\stackrel{<}{\scriptstyle\sim}}
\newcommand{\gsimeq}{\stackrel{>}{\scriptstyle\sim}}
\newcommand{\mayany}{\stackrel{>}{\scriptstyle<}}

\renewcommand{\r}{{\bf r}}
\newcommand{\deltak}{\ifmmode{\left\{ \delta , K \right\}}
\else{$\left\{ \delta, K \right\}$ }\fi}
\def\ambec{AMBEC\noeq{eq:gpe}}
\newcommand{\commentout}[1]{{}}

\begin{document}

\title{%
  Local eigenstates and parametric excitations of a simple
  atom-molecule Bose-Einstein condensate
}
\author{Marijan Ko\v{s}trun and Juha Javanainen}
\affiliation{%
  Department of Physics, University of Connecticut, Storrs,
  Connecticut 06269-3046
}
\email{kostrun@phys.uconn.edu}
\date{\today}

\begin{abstract}
  We model an atom-molecule Bose-Einstein condensate (AMBEC)
  using simplified set of coupled Gross-Pitaevskii equations (GPE),
  where we neglect the background (elastic) scattering length of the atoms.
  We analyze the ground state numerically and analytically, and construct its
  twin state through transformation $\delta\rightarrow-\delta, K\rightarrow-K$.
  We find that the ground state is a collection of three local eigenstates:
  all-atom state, mixed (atom-molecule) state, and all-molecule state,
  while the twin state comprises of mixed state with tunable fraction of atoms including unity
  but excluding all-molecules.
  We find the analytic boundaries of the local eigenstates from
  the stability analysis of the underlying all-molecule and all-atom
  eigenstate.
  In the ground state we find either regular oscillations in size and fraction of both condensates,
  or shrinking of the atomic condensate that resembles the collapse.
  In the twin state we find rapid irregular disintegration of both condensates.

  We contrast the properties of the mean-field parametric excitations found in AMBEC with
  the experimentally observed excitations of the BEC.
  We discuss some enticing possibilities for creation of atomic BEC of controllable
  half-width by performing the parameter sweep around the boundary.
\end{abstract}


\maketitle

\section{Introduction}

Bose-Einstein condensation (BEC) in weakly interacting atomic gases
continues to be of considerable experimental and theoretical interest.
BEC dynamics strongly depends on atom-atom interactions,
which is quite accurately described by their s-wave scattering length $a$.
It was argued~\cite{Dalfovoetal1996,Duineetal2000,Gertonetal2000,Donleyetal2001:bosenova,DuineStoof2003:bosenova}
that BEC with $a<0$ may collapse in finite time~\cite{Wadatietal1998}
resulting in a gas-liquid phase transition~\cite{Stoof1994:phasetransition,Timmermansetal1999B,glc}.
Controlling the collapse is one way of controlling the
atom-atom interactions~\cite{Cornishetal2000,Donleyetal2001:bosenova,Kokkelmansetal2002,Grimmetal2003:BECinCs}.
As is known, atom-atom interactions can be manipulated
through Feshbach resonances in magnetic field
or through photo-association~\cite{Fedichevetal1996:nrl,BohnJulienne1999,Kostrunetal2000:pa,Robertsetal2001:tunableBEC}.
Either mechanism can be described as
a coupling of atomic pair to a molecular dimer~\cite{Drummondetal1998,jj1998:pa1,Kaplanetal1996:eftscattering}.

The dynamics of dilute atomic BEC can be described using the mean-field
theory~\cite{ParkinsWalls1998},
in which the macroscopic condensate wave function $\Phi=\Phi( {\bf r} )$
is a solution of a Gross-Pitaevskii equation (GPE),
\begin{equation}
  \label{eq:gpe:1}
  \begin{array}{rcl}
    i \hbar \frac {\partial \Phi( {\bf r}, t)}{\partial t} &=&
    \left( - \frac {\hbar^2} {2 m} \nabla^2 + V_{0}({\bf r},t) \right)
    \Phi( {\bf r}, t)\\
    &+& \frac{4\,\pi\,\hbar^2\,a\,N}{m}
    \left| \Phi ({\bf r},t) \right|^2 \Phi( {\bf r}, t).
  \end{array}
\end{equation}
Here, we choose $\Phi$ normalized to unity with $N$ being the number of atoms in the
condensate, $V_0$ the trapping potential, while $m$ is the mass of an atom
and $a > 0$ atom-atom s-wave scattering length.

It has been recognized that the mean-field approach of \eq{eq:gpe:1}
may fail in certain situations, however, the cause of failure
is embedded in the mean-field equations themselves:
E.g., so-called collapse of the condensate for $a \lsimeq -0.67$ is
a parametric excitation of \eq{eq:gpe:1} in which the (non-linear)
eigenstate present and continuous for $a \gsimeq -0.67$ disappears.
One can thus conjecture for single-species BEC that the mean-field
approximation is likely to fail if it predicts parametric excitations
of some sort.
A renormalization of negative scattering length calls for coupled
atomic and molecular dimer BEC, so their mean-field
theory needs to be examined with respect to parametric excitations.

We build phenomenological mean-field theory of atom-molecule BEC (AMBEC)
from three parameters: $K$, coupling strength between a pair of atoms and a molecule, $\delta$,
the detuning of the molecular state\footnote{%
  Here the detuning $\delta$ is given by
  $\delta = \nu_1 - E_{b}/\hbar$, where $\nu_1$ is the laser
  frequency and $E_{b}$ the energy of the molecular state$E_{b}$.
  We choose this convention so that the sign of the effective atom-atom
  scattering length conveniently coincides with the sign of detuning.
},
and the background atom-atom scattering length~\cite{Timmermansetal1999}.
However, close to the resonance the atom-atom scattering length is
dominated by the resonant contribution so there the background scattering
length can be safely ignored.
Their Gross-Pitaevskii equations read,
\begin{subequations}
  \label{eq:gpe}
  \begin{eqnarray}
    i \frac {\partial \varphi} {\partial \tau} & = & H_{a} \varphi -
    K \varphi^* \psi,
    \label{eq:gpe_phi}
    \\
    i \frac {\partial \psi} {\partial \tau} & = &
    \left( H_{m} - \delta \right) \psi - K \varphi^2.
    \label{eq:gpe_psi}
  \end{eqnarray}
\end{subequations}
The Hamiltonians $H_{a}$ and $H_{m}$ correspond to the center-of-mass
energies for atoms and molecules, respectively,
\begin{subequations}
  \begin{eqnarray}
    H_{a} & = & -\half {\nabla^2}
    + \half {\r}^2, \\
    H_{m} & = & -\frac 1 4 {\nabla^2}
    + {\r}^2\,.
  \end{eqnarray}
  \label{eq:ho}
\end{subequations}%
For simplicity, we assume that both atomic and molecular trap are
isotropic with the same trap frequency $\omega$, which we then
use to scale all the other frequencies.
Similarly, we use characteristic length of the atomic trap
$L_0 = \sqrt{\frac{\hbar}{m\omega}}$ to scale all the distances.
Lastly, we scale the fields with respect to the total number of particles,
so that the normalization reads
\begin{equation}
  \label{eq:normalization}
  \left<\varphi | \varphi \right> + \left< \psi | \psi \right> = 1.
\end{equation}
The expression for the conserved total energy of the system is then
\begin{equation}
  \label{eq:energy}
  E = \left< \varphi | H_{a} | \varphi \right> +
  \half \left< \psi | H_{m} - \delta | \psi \right>
  + E_{I},
\end{equation}\
where
\begin{equation}
  \label{eq:pa interaction}
  E_{I} = - \half K \left(
    \langle \varphi^2 | \psi \rangle +
    \langle \psi | \varphi^2 \rangle
  \right).
\end{equation}

In this paper we examine two collections of eigenstates of AMBEC~\noeq{eq:gpe}
that are symmetric with respect to the resonant detuning $\delta$:
the ground state and its ``twin state.''
We use zero-dimensional analysis to find the local eigenstates in
parameter space, and examine their stability.
We then extend analysis to the full mean-field AMBEC equations numerically and analytically.
We find the extent of local eigenstates comprising each collection, and examine
$(i)$, their parametric excitations; and $(ii)$, how in certain parametric sweeps
the same excitations can be avoided.

\section{Zero-dimensional AMBEC }

For zero-dimensional analysis,
we assume free condensates,
which are described by plane waves of varying amplitudes~\cite{Timmermansetal1999,Kostrunetal2000:pa,WuCote2002}.
We absorb all the constants in $\delta \leftarrow \delta/K$,
so the atomic $\alpha = \alpha(t)$ and the molecular $\beta=\beta(t)$
amplitude are the solutions of
\begin{subequations}
\label{eq:0mode:ambec}
\begin{eqnarray}
  i\dot\alpha & = & - \alpha^* \beta,\\
  i\dot\beta  & = & -\delta\,\beta - \alpha^2.
\end{eqnarray}
\end{subequations}
We expand time dependency of the amplitudes as
$\alpha = x \cdot \exp(-i\,\mu\,t)$ and
$\beta = y \cdot \exp(-i\,2\,\mu\,t)$,
with $x$ and $y$ real, and normalize them so that $x^2 + y^2 \equiv 1$.

\begin{figure}[hp]
  \centering
  \includegraphics[type=pdf,ext=.pdf,read=.pdf,scale=0.8,clip]{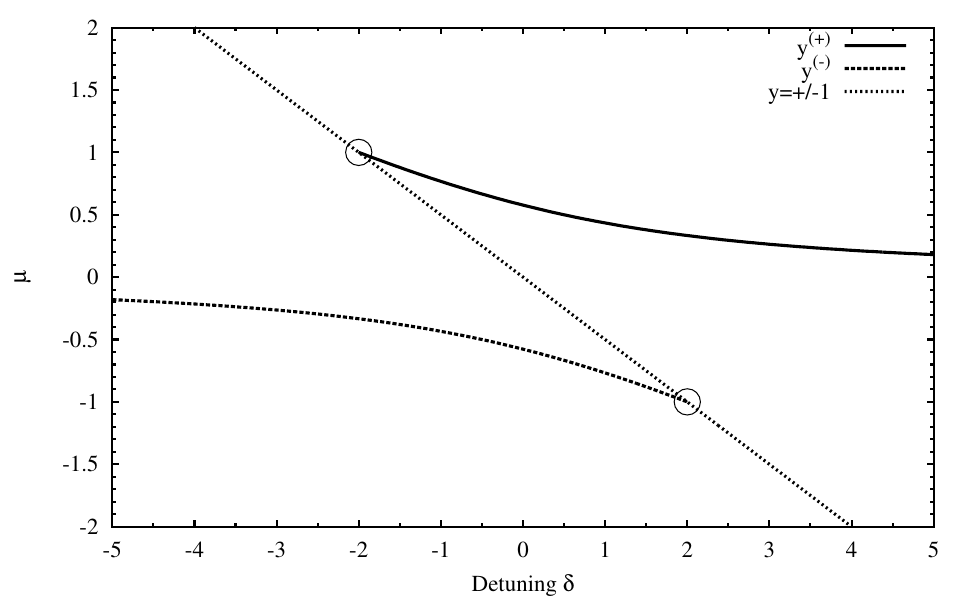}
  \vskip1em
  \caption{%
    Energy $\mu$ for the stationary states of \eq{eq:0mode:ambec},
    with the critical points $\delta=\pm2$.
    \label{fig:2m:mu}
}
\end{figure}

We find the stationary solutions of \eq{eq:0mode:ambec} as:
\begin{itemize}
\item
  The all-molecule state:
  \begin{equation}
  y \equiv \pm1,
  \end{equation}
  which exists for $\forall \delta$, and with $\mu = -\delta/2$.
\item
  The mixed atom-molecule states:
  \begin{equation}
  \label{eq:0mode:yminus}
  y^{(-)}( \delta ) = \frac{\delta}{6}  + \frac{\sqrt{\delta^2 + 12}}{6},
  \end{equation}
  which exists for $\delta \le 2$, and
  \begin{equation}
  \label{eq:0mode:yplus}
  y^{(+)}( \delta ) =
  \frac{\delta}{6} - \frac{\sqrt{\delta^2 + 12}}{6},
  \end{equation}
  which exists for $\delta \ge -2$.
  \\
  For both mixed states we have $\mu^{(\pm)} = -y^{(\pm)}$.
\end{itemize}
In~\fig{fig:2m:mu} we show $\mu = \mu(\delta)$ for all stationary solutions.

To examine their stability we rewrite \eq{eq:0mode:ambec}
in terms of the positive molecular amplitude $y = |\beta|$,
and the phase difference $\theta = 2 \arg(\alpha) - \arg(\beta)$,
\begin{subequations}
\label{eq:0mode:ytheta}
\begin{eqnarray}
\dot y      & = & (1-y^2) \sin \theta,\label{eq:0mode:y}\\
\dot\theta  & = & \delta +
\left( \frac 1 y - 3 y \right) \cos\theta \label{eq:0mode:theta}.
\end{eqnarray}
\end{subequations}
The stationary solutions have $\theta=0$ or $\pi$.
The all-molecule state $y\equiv1$ can have either $\theta$.
The two mixed atom-molecule states are given
for $\theta=0$ by $y^{(-)}$ from \eq{eq:0mode:yminus},
and for $\theta=\pi$ by $-y^{(+)}$ from \eq{eq:0mode:yplus}.
We find the frequency of small oscillations $\lambda$ near the stationary states $y^{(\pm)}$ as,
\begin{equation}
\label{eq:0mode:freq}
\lambda^2 = \cos^2 \theta \, (1-y^2) \, (\frac{1}{y^2}+3),
\end{equation}
which is always non-negative. We thus conclude that the mixed states are stable.
\\
The all-molecule state we find stable everywhere except for $|\delta| < 2$, cf.~\cite{WuCote2002}.
In the limit $|\delta| \gg 1$ we find $\theta \approx \delta \cdot t$,
while $\delta y(t) = 1 -y(t)$ behaves as,
\begin{equation}
  \label{eq:0mode:allmolecule}
  \delta y(t) \approx \delta y (0) \cdot \exp\left[
  -\frac{\sqrt2}{\delta} ( 1- \cos(\delta\cdot t))
\right].
\end{equation}
Thus, in the limit $t\rightarrow\infty$ small oscillations between atoms and molecules
persist and do not die out.

From the all-molecule and the mixed states we construct the ground state
around $\theta=0$, and which exists for $\forall\delta$:
$y_{GS} = y^{(-)}$ for $\delta < 2$, and $y_{GS}=1$ for $\delta > 2$.
We construct the second eigenstate around $\theta=\pi$, which
we call the twin state:
$y_{TW} = y^{(+)}$ for $\delta > -2$, and $y_{TW}=-1$ for $\delta < -2$.
One can see that the twin state is symmetric to the ground state with respect
to $\delta \rightarrow -\delta$ and $y \rightarrow -y$.
We remark that $\delta = \pm2$ are the critical points where the mixed state
coincides with the all-molecule state but the transition is non-smooth:
In parameter sweeps these points act as a source of parametric excitations.

\section{Numerical Investigation}

We turn to numerical methods~\cite{Kostrun2001:DS} to find the AMBEC~\noeq{eq:gpe}
ground state and examine the properties of the underlying stationary states,
the ground state and the twin state, by parameter sweeps in $K$ and $\delta$ directions.
For compactness of presentation, in our calculations we limit ourselves
to spherically symmetric trap, that is, the atomic and molecular Hamiltonians
contain only radial operators.

\subsection{Ground State}

We find the ground state through numerical iteration in complex time, followed by renormalization.
We focus on the integer mesh $\delta = -30 \ldots 30$, and $K = 0.1,1 \ldots 30$,
where we use $K=0.1$ as an approximation of the limit $K\rightarrow0$.
For each pair $(\delta,K)$ we find the condensate wave
functions of atoms $\varphi$, and of molecules $\psi$.
We present features of the atomic condensate in terms of its fraction $N_a = \braket\varphi\varphi$,
and the half-size $R_{1/2}$,
\begin{equation}
\label{eq:app:halfsize}
\frac
{ \int^{R_{1/2}}_0\, dr\, r^2\, | \varphi(r)|^2 }
{ \int^{\infty}_0 \, dr \, r^2 \, | \varphi(r)|^2 }
= {1\over2}.
\end{equation}

\begin{figure}[p]
  \centering
  \includegraphics[type=pdf,ext=.pdf,read=.pdf,scale=1,clip]{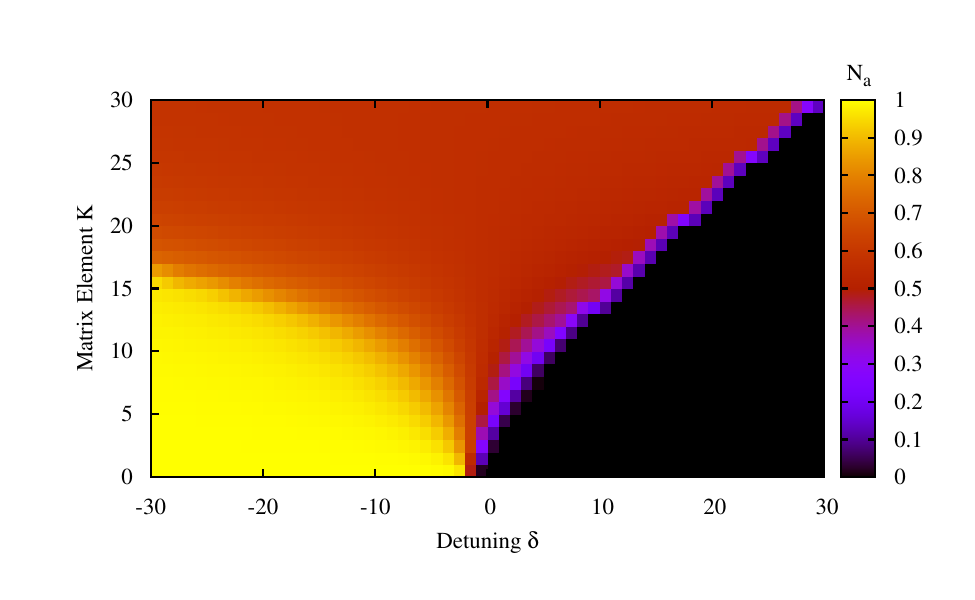}
  \vskip1em
  \caption{%
    (Ground state) Fraction of atoms $(N_a)$ as a function of $\delta$ and $K$ we find
    from evolution of GPEs~\noeq{eq:gpe} in complex time.
    One can easily recognize the regions of the mostly-atom state (in yellow, for $\delta < 0$) and
    the all-molecule state (in black, for $\delta > 0$).
    \label{fig:gs:na}
}
\end{figure}

$N_a$ and $R_{1/2}$ calculated for the ground state that
we show in~\figs{fig:gs:na}{fig:gs:cra}
strongly suggest that the ground state comprises three local eigen-states:
mostly-atom state, mixed atom-molecule state, and the all-molecule state.
We surmise the existence of two boundaries,
one $f^1_{GS}$, between the mostly-atom and mixed state,
and the other $f^2_{GS}$, between the all-molecules and the mixed state.

\begin{figure}[p]
  \centering
  \includegraphics[type=pdf,ext=.pdf,read=.pdf,scale=0.5,clip]{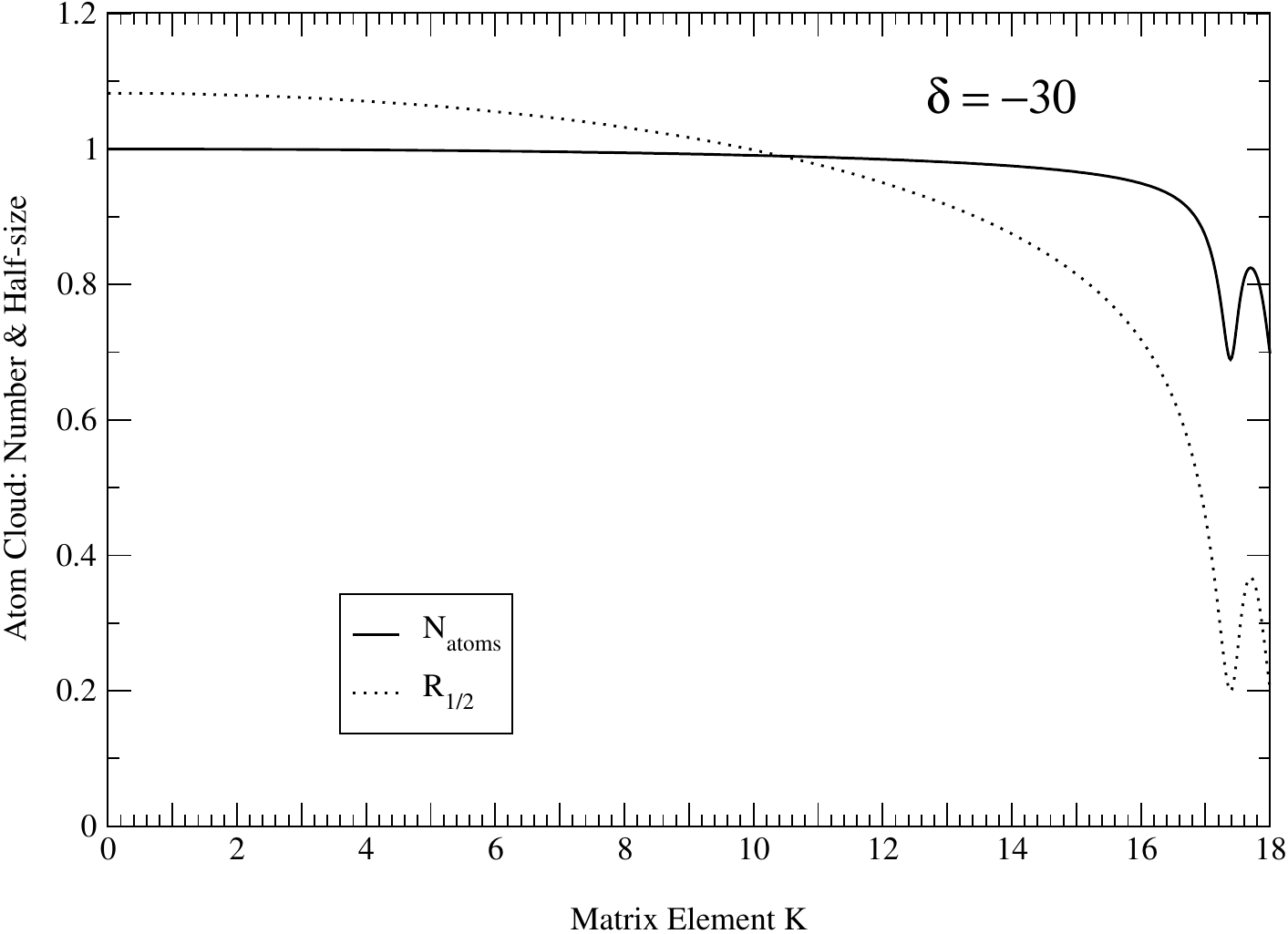}
  \caption{%
    (Ground state) $N_a$ and $R_{1/2}$ in parameter sweep
    in $K=0\ldots18$ with fixed $\delta=-30$.
    Oscillations near $K = 17$ suggest a parametric excitation associated with crossing
    the boundary $f^1_{GS}$.
    \label{fig:gs:belowresonanceintensitysweep}
}
\end{figure}

\begin{figure}[tp]
  \includegraphics[type=pdf,ext=.pdf,read=.pdf,scale=0.5,clip]{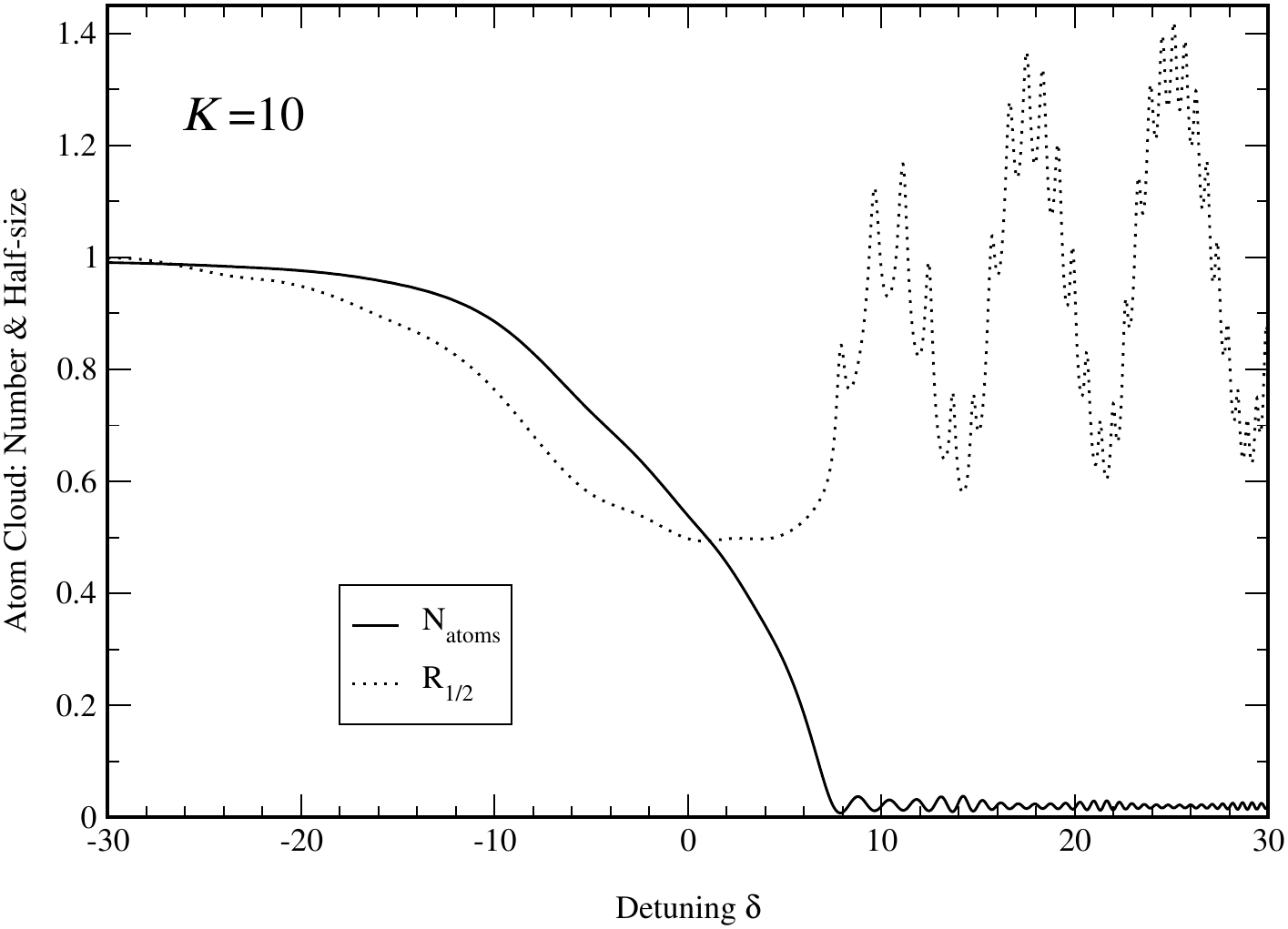}
  \caption{%
    (Ground state) $N_a$ and $R_{1/2}$ in parameter sweep
    of $\delta = -30 \ldots 30$ with $K=10$ fixed.
    Oscillations near $\delta=8$ suggest a parametric excitation from crossing
    the boundary $f^2_{GS}$.
    Curiously, the anticipated oscillations from $f^1_{GS}$ near $\delta=0$ are absent.
    \label{fig:gs:positivedetuningsweep}
}
\end{figure}

We confirm presence of the boundaries $f^1_{GS}$ and $f^2_{GS}$ through
parametric sweeps: They act as a source of parametric excitations.
\\
Firstly, in~\fig{fig:gs:belowresonanceintensitysweep}
we show that the sweep of $K = 0 \ldots 18$ for fixed $\delta=-30$
reveals a parametric excitation near $K=17$, which we associate with $f^1_{GS}$.
\\
Secondly, in~\fig{fig:gs:positivedetuningsweep} we show
that the sweep of $\delta=-30\ldots30$ for fixed $K=10$
reveals a parametric excitation near $\delta=8$, which we associate with $f^2_{GS}$.
Interestingly, for $\delta<0$ we do not see the evidence of $f^1_{GS}$.

We address the differences between excitations at $f^1_{GS}$ and $f^2_{GS}$
in Sec. V, when we present our analytic treatment of the boundaries.

\subsection{Twin State}

\begin{figure}[tp]
  \centering
  \includegraphics[type=pdf,ext=.pdf,read=.pdf,scale=0.5,clip]{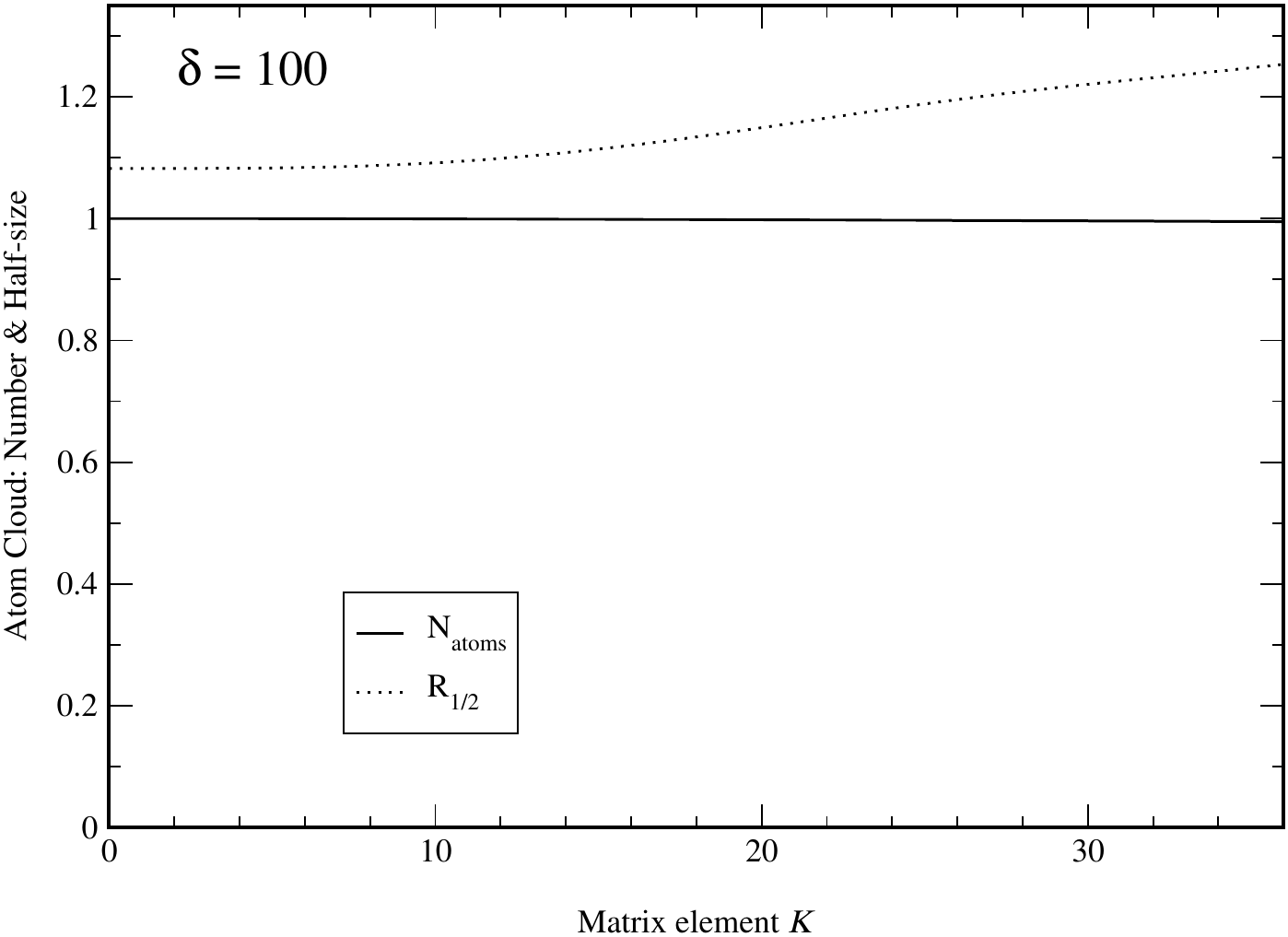}
  \caption{%
  (Twin state) Parametric sweep of $K=0 \ldots 36$ with fixed $\delta=100$
  of an all-atom state shows no evidence of parametric excitations.
  We use this technique to create the twin state at desired $K$ and $\delta>0$.
  \label{fig:tw:aboveresonanceintsweep}
}
\end{figure}

Based on the zero-dimensional model,
the twin state is an all-atom state for $\delta>0$.
We create it by adiabatically evolving an all-atom state found for $\delta>0$ and$K\equiv0$,
to desired value of $K$.
We demonstrate the effectiveness of this technique in~\fig{fig:tw:aboveresonanceintsweep},
where we start from an all-atom state and sweep $K=0\ldots36$ for fixed $\delta=100$.
This also demonstrates that the twin state is not fully symmetric with the
ground state as an equivalent of the boundary $f_{GS}^1$ that would be constructed
by changing the sign of $\delta$, is missing.

\begin{figure}[tp]
  \centering
  \includegraphics[type=pdf,ext=.pdf,read=.pdf,scale=0.5,clip]{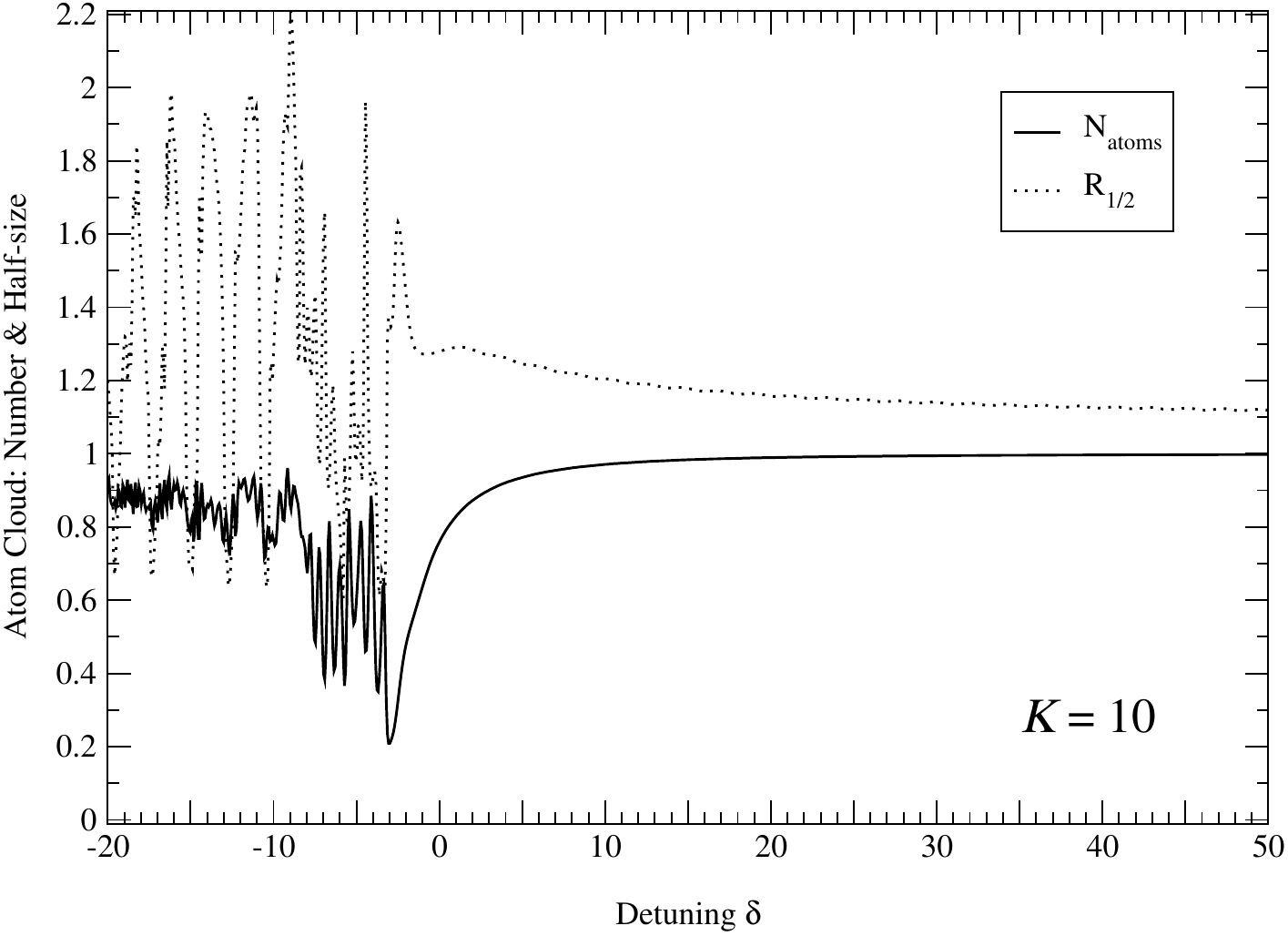}
  \caption{%
    Parametric sweep of all-atom state
    for $\delta = 50 \ldots -30$
    with $K=10$ fixed reveals parametric excitation near $\delta=-3$.
    \label{fig:tw:negativedetuningsweep}
}
\end{figure}

However,
in parameter sweep in $\delta = 50\ldots -30$ for fixed $K=10$,
which we show in~\fig{fig:tw:negativedetuningsweep},
we do find parametric excitations near $\delta = -3$.
This suggests presence of the boundary $f_{TW}$, which is analogous to $f^2_{GS}$.

\section{Analytic Treatment of the Boundaries}

\subsection{%
  Boundary $f^1_{GS}$ and the extent of the all-atom state \label{sec:allatoms}
}

We start by recalling that in the limit $K\rightarrow0$ and $-\delta \gg K$,
we can write the approximate solution of \ambec\ for mean-field $\psi$ as,
$\psi \simeq -\frac{K}{\delta} \varphi^2$.
This then appears in the atomic GPE~\noeq{eq:gpe:1}
as a tunable atom-atom s-wave scattering length
$a_{f} =\frac{K^2}{4\,\pi\,\delta} < 0$.
In the literature, it is known that atomic BEC can exist with small negative $a_{f}$.
When this $a_f$ is manipulated (increased in absolute value)
to the critical value $a_{f} \rightarrow -0.67$ the atomic BEC collapses.

The expression for the effective scattering length $a_f$ above
is singular near $\delta=0$.
We now show how singularities can be removed under the assumption that
both mean-fields are stationary.
We start by writing the equation for $\psi$ in AMBEC,
\begin{equation}
  \label{eq:psi}
    \left(
      i{\partial\over\partial t} + \delta - H_{m} \right) \, \psi = -K \varphi^2,
\end{equation}
In the limit in which the fraction of the atoms is much greater than
the fraction of the molecules, $\braket{\varphi}{\varphi} \gg \braket{\psi}{\psi}$,
the evolution of $\psi$ is completely dominated by $\varphi^{2}$.
We assume that the eigen-energy is $\epsilon$, so that the time dependence is $\varphi \sim e^{-i \, \epsilon\, t}$
and $\psi \sim  e^{-i \, 2 \, \epsilon\, t}$, and
that we can neglect other terms in expansion of $\psi$,
that is, $H_m \, \psi \approx E_{0,m}\,\psi$,
where $E_{0,m} = \bra{\psi}H_m\ket{\psi} / \braket{\psi}{\psi}$.
Then, we can solve \eq{eq:psi} for the mean-field $\psi$ as,
\begin{equation}
  \label{eq:psi:sol}
  \psi \approx - \frac{K}{2\,\epsilon + \delta - E_{0,m}} \, \varphi^{2}.
\end{equation}
We can now use the atomic part of \ambec\ to write the equation for $\epsilon$ as,
\begin{equation}
  \label{eq:abec:ea}
  \epsilon = E_{0,a} + \frac{K^{2}}{2\,\epsilon + \delta - E_{0,m}} \cdot
    \frac{\braket{\varphi^{2}}{\varphi^{2}}}{\braket{\varphi}{\varphi}}.
\end{equation}
Here we use $H_a \, \varphi \approx E_{0,a} \varphi$, where
$E_{0,a} = \bra{\varphi}H_a\ket{\varphi} / \braket{\varphi}{\varphi}$,
similarly to what we have done for the molecules.
The two solutions of \eq{eq:abec:ea} are $\epsilon^{\pm}$, where
\begin{equation}
  \label{eq:abec:ea:sol}
  \epsilon^{\pm} = \frac14\,\left(2\, E_{0,a} + E_{0,m}-\delta\right) \pm
  \frac14\sqrt{\left(2\, E_{0,a} - E_{0,m}+\delta\right)^{2} +
    8\,K^2\,\frac{\braket{\varphi^{2}}{\varphi^{2}}}{\braket{\varphi}{\varphi}}}.
\end{equation}
where the negative branch corresponds to the ground state,
while the positive branch to the twin state.
We remark that $\epsilon^{\pm}$ in \eq{eq:abec:ea:sol} are everywhere well
behaved, and that $\epsilon^{\pm}$ differ in sign of $2\,\epsilon + \delta - E_{0,m}$.

We use variational technique~\cite{Timmermansetal1999,Dalfovoetal1999,ParkinsWalls1998}
to find $\epsilon = \epsilon(s)$ for the all-atom state, $\braket{\varphi}{\varphi}=1$,
where $\varphi$ is given by
\begin{equation}
  \label{eq:allatomgaussian}
  \varphi(r;s) =
  \frac{1}{ \left( \pi\,s^2\right)^{3/4}}
  e^{-\frac{r^2}{2\,{s^2}}}.
\end{equation}
The variational energies are
\begin{subequations}
\begin{equation}
 E_{0,a} = \frac{3\,(1+s^{4})}{4\,s^{2}} = E_{0,m},
\end{equation}
\begin{equation}
 \braket{\varphi^{2}}{\varphi^{2}} = \frac{1}{2\,\sqrt2\, \pi^{3/2} \, s^{3}},
\end{equation}
\end{subequations}
which then enter the expression for $\epsilon$, \eq{eq:abec:ea:sol}.
We find roots of $\partial \epsilon/\partial s = 0$ that
are real and positive, and determine
which of them are local minima of $\epsilon(s)$.
We keep in mind that $s \gsimeq 1$ is a signature of an all-atom state,
while $s \ll 1$ of the mixed- or all-molecule state.

\begin{figure}[tp]
  \centering
  \includegraphics[type=pdf,ext=.pdf,read=.pdf,scale=1,clip]{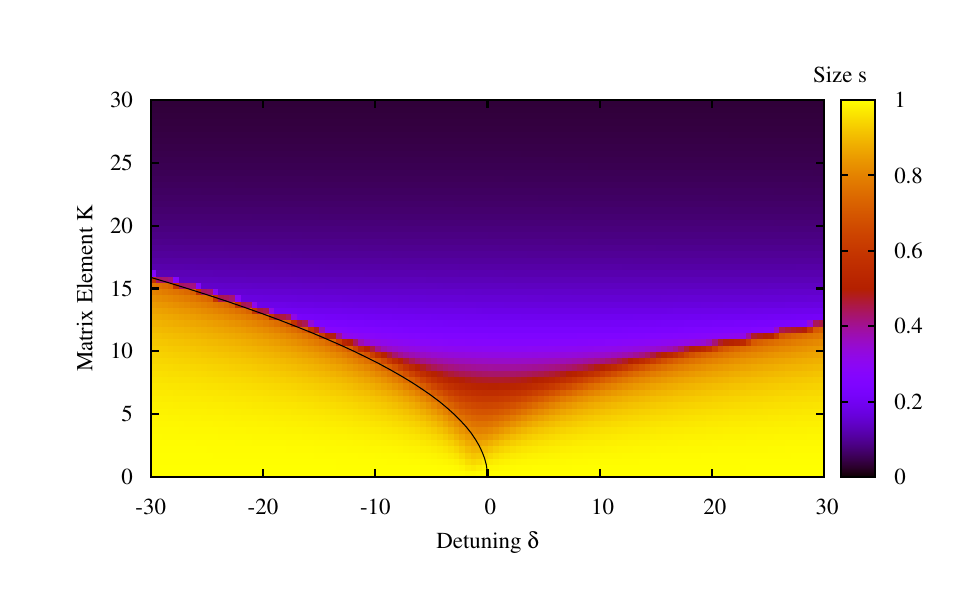}
  \caption{%
    The variational size of the atomic BEC in AMBEC as a function of detuning
    $\delta$ and matrix element $K$.
    For comparison the solid black line is the extent of the ground state in the pure
    atomic BEC, $K^{2}/(4\,\pi\,\delta) = -0.67$,
    and it coincide almost everywhere with $f^1_{GS}$ except near origin,
    where $f^1_{GS}$ disappears.
    \label{fig:f1gs:var}
}
\end{figure}

In \fig{fig:f1gs:var} we show the variational size $s$ as a function
of $\delta$ and $K$ for the negative branch of $\epsilon$, which corresponds
to the ground state.
We see that for larger negative $\delta$'s the boundary $f^1_{GS}$
of the all-atom state coincide with the expected
$\frac{K^2}{4\,\pi\,\delta} \simeq -0.67$.
However, for $K \lsimeq 15$ the boundary disappears near $\delta\sim-10$,
just as we have seen in the numerical simulations,
cf.~\fig{fig:gs:positivedetuningsweep}.
The reason why is because by approaching the expected boundary
the fraction of atoms $N_a$ begins to decrease,
and this modulates the coupling as $K \cdot N_a$.
This effect is so small, however, that it is accessible only for small initial
magnitudes of coupling $K$.

An equivalent analysis for the twin state is utterly uninsightful: continuous variational solution exists
in the entire parameter space.
We thus turn our attention to the opposite limit - that of the all-molecule state.

\subsection{Boundaries $f^2_{GS}$ and $f_{TW}$ of the all-molecule state}

\subsubsection{Existence of variational solutions for the atoms}
\label{sec:existence}

\begin{figure}[tp]
  \centering
  \includegraphics[type=pdf,ext=.pdf,read=.pdf,scale=0.5,clip]{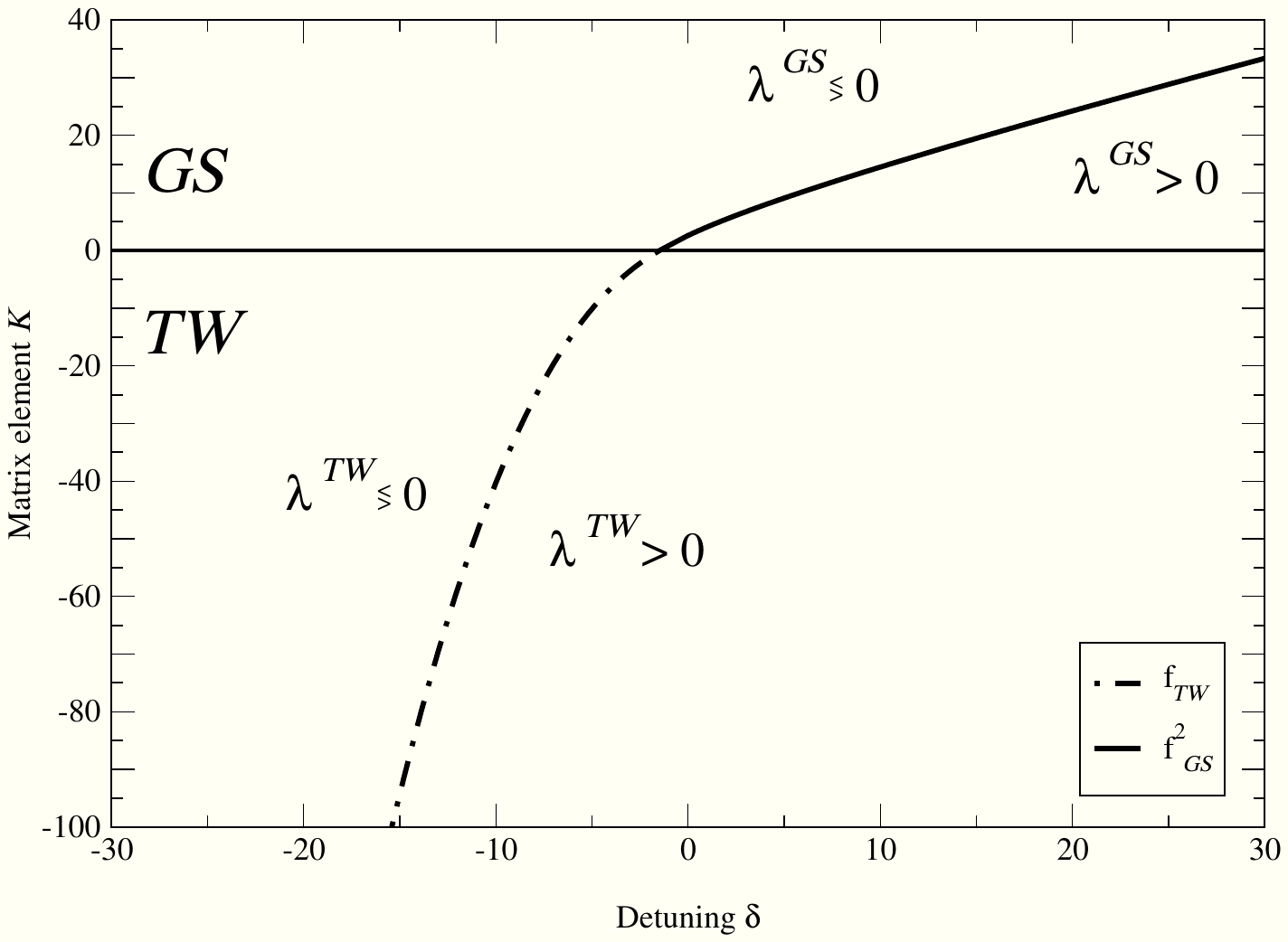}
  \caption{%
    Variational infinitesimal atomic configurations
    may exist only if \eq{eq:atomiceigenstatevariation} has real roots,
    which we write as $\lambda^{TW,GS}\mayany0$.
    As discussed in text, the case $K>0$ pertains to the ground state,
    while $K<0$ to the twin state.
    For $\lambda^{TW,GS} > 0$ there are no variational atomic configurations.
    The boundaries $f_{TW}$ and $f_{GS}^2$ separate the regions with atomic configurations
    from the regions without.
    \label{fig:allmolecule}
}
\end{figure}

We examine stability of the all-molecule ground state in the
presence of infinitesimal atomic configurations.
The all-molecule state is the ground state of $H_m$,
\begin{equation}
  \label{eq:allmoleculeswfn}
  \psi(r) =
  \left( \frac{2}{\pi} \right)^{3/4}
  e^{- {r^2}}.
\end{equation}
Its eigen-energy is easily found in the absence of atomic mean-field,
\begin{equation}
  \label{eq:allmoleculeeigenenergy}
  \mu = {1\over2}
  \left( \frac 3 2 - \delta \right).
\end{equation}
The atomic mean-field small in fraction, $\braket{\varphi}{\varphi} \ll 1$,
is then described by,
\begin{equation}
  \label{eq:atomiceigenstate}
  \mu \varphi =
  H_a \varphi -
  K \varphi^* \psi,
\end{equation}
where $\mu$ is given by \eq{eq:allmoleculeeigenenergy}.
We now approximate the atomic mean-field with \eq{eq:allatomgaussian},
and define difference $\lambda^{GS}(s)$, where
\begin{equation}
  \label{eq:lambdags}
  \lambda^{GS}(s) \, \braket{\varphi}{\varphi} = \bra\varphi H_a \ket\varphi - \mu
  -K \braket{\varphi^2}\psi.
\end{equation}
We notice that for (real) $\varphi$ the atomic norm cancels out.
Valid sizes for the atomic condensates $s$ satisfy,
\begin{equation}
  \label{eq:atomiceigenstatevariation}
  \lambda^{GS}(s) = 0.
\end{equation}
We find that \eq{eq:atomiceigenstatevariation} has either
two solutions for $s$ (we write this as $\lambda^{GS} \mayany 0$, that is,
$\lambda_{GS}$ is positive for some values of $s$ and negative for the others),
or has no solution ($\lambda^{GS}(s) > 0, \forall s$).
We identify as $f^2_{GS}$ the boundary between the two regions,
for which $s$ is a double root.
In~\fig{fig:allmolecule} we show the boundary $f^2_{GS}$ in the parameter space.

For the twin state, the phase shift between the molecular
amplitude $\psi$ and the matrix element $K$ is $-1$,
so the new difference $\lambda^{TW}$ now reads,
\begin{equation}
    \lambda^{TW} \, \braket{\varphi}{\varphi}=
    {\bra\varphi} H_a \ket\varphi - \mu + K \braket{\varphi^2}\psi.
    \label{eq:lambdatw}
\end{equation}
Again, valid sizes for the atomic condensates $s$ satisfy,
\begin{equation}
  \label{eq:atomiceigenstatevariation:twin}
  \lambda^{TW}(s) = 0.
\end{equation}
The rest of the analysis is identical to what we have done for the ground state.
In \fig{fig:allmolecule} we combine the findings for the ground state ($K>0$) and
the twin state ($K<0$).
Here, we identify the boundary as $f_{TW}$.

In \tab{tab:f2gstheoryvsnumericks} we give $f^2_{GS}$
as $K$ for various detunings $\delta$,
where $K_{num}$ are the numerical values we find by solving the full AMBEC,
while $K_{var}$ are the variational solutions.
For comparison, we also give numerical values for the fraction ($N_a^{mix}$) and the
half-size ($R_{1/2}^{mix}$) of the atomic distribution on the mixed side
(on the atomic side both values are near unity).
From \tab{tab:f2gstheoryvsnumericks} we see that across $f^2_{GS}$
the size and the fraction of atomic condensate change discontinuously.
The boundary thus acts as a source of parametric excitations for parameter sweeps
crossing it.

We numerically examine $f_{TW}$ through parameter sweeps of the twin state for $K=10, 20$.
In \tab{tab:ftwtheory} we list the detunings at which the condensates disintegrate
as $\delta_{num}$ and the fraction of atoms at the onset of the instability
as $N_a^{mix}$.
We see that the mixed state disintegrates very close to $\delta = 0$.
In this region the effective scattering length is large positive number.
The analysis of $\epsilon^+$ suggests that at $f_{TW}$ the variational size $s$ of the atomic condensate
is maximal.

We show how these traits are related to the stability of the
all-molecule state.

\subsubsection{Stability of the all-molecule state}

We write the solution of \eq{eq:gpe_phi} in the form
\begin{equation}
  \label{eq:allatomsolution}
  \varphi(\tau) =
  \left( a(\tau) + i\,b(\tau) \right) \, \varphi(r;s) \, \exp(-i\mu\tau),
\end{equation}
where $\varphi(r;s)$ is the unit-normalized variational atomic amplitude,
\eq{eq:allatomgaussian},  and $\mu$ is the
eigen-energy of the all-molecule state, \eq{eq:allmoleculeeigenenergy}.
The time dependent real functions satisfy
$|a(\tau)| \ll 1$ and $|b(\tau)| \ll 1$.
We integrate out the eigenstate $\varphi$ to yield a system
of ordinary differential equations,
\begin{equation}
  \label{eq:ode:ab}
  \begin{array}{rcrl}
    \dot a & = &
    \left(
      \bra\varphi H_a \ket\varphi - \mu
      +K \,\braket{\varphi^2}{\psi}
    \right) & b, \\
    \dot b & = &
    -\left(
      \bra\varphi H_a \ket\varphi - \mu
      -K\,\braket{\varphi^2}{\psi}
    \right)
    & a.
  \end{array}
\end{equation}
The frequency $\lambda$ of small excitations is given by,
\begin{equation}
\label{eq:ode:eigenfreq}
\begin{array}{rcl}
  \lambda^2(s) &=& \left( \bra\varphi H_a \ket\varphi - \mu \right)^2
  - {K^2} |\braket{\varphi^2}{\psi}|^2\\
  & = & \lambda^{GS}(s) \cdot \lambda^{TW}(s).
\end{array}
\end{equation}
Stability of the molecular mean-field $\psi$ then requires that
$\lambda^2(s)>0, \forall s>0$.
Conversely, if $\exists s'>0$ such that $\lambda^2(s')<0$, then this particular
atomic mean-field may grow in fraction exponentially,
and so destroy the ``all-molecule'' character of the eigenstate.
Interestingly, this also means that if the molecular mean-field is
stable ($\lambda^2>0$)
then there are no available variational solutions for the
atomic mean-field (it requires $\lambda^{GS} = 0$ or $\lambda^{TW} = 0$),
and \viceversa.
Unavailability of of the atomic configurations in the vicinity of all-molecule
state has been hinted by the zero-dimensional AMBEC,
where the same occurs, cf.~\eq{eq:0mode:allmolecule}

\begin{figure}[htpb]
  \centering
  \includegraphics[type=pdf,ext=.pdf,read=.pdf,scale=0.5,clip]{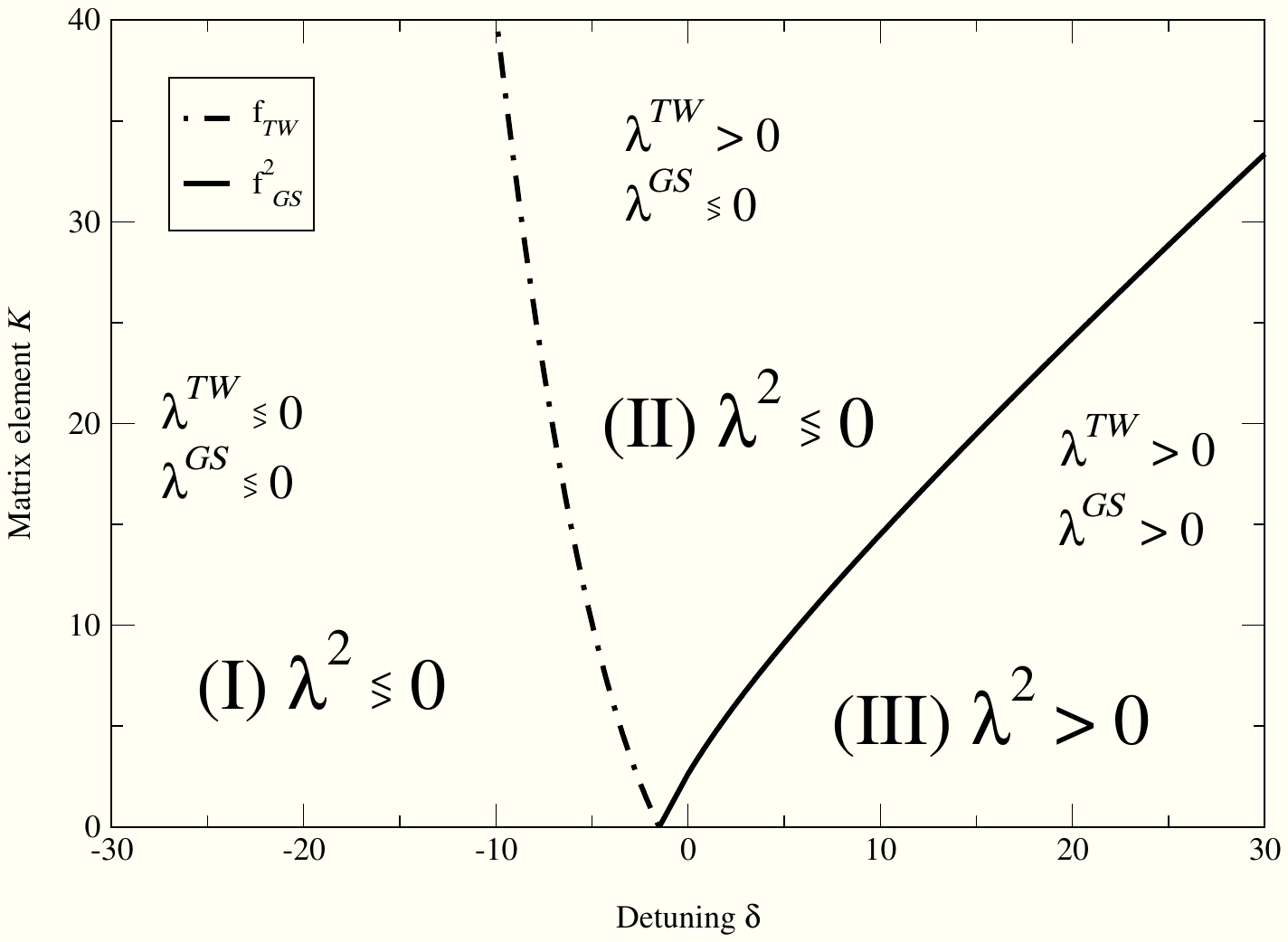}
  \caption{%
    Regions of $\lambda^2$, \eq{eq:ode:eigenfreq}, in the
    parameter space. $\lambda^2$ is positive only in Region III,
    bounded by $f_{GS}^2$, for all variational atomic sizes $s$.
    In all other regions there exist unstable variational solutions
    for $s$, for which $\lambda^2<0$.
    \label{fig:allmolecule:o2}
}
\end{figure}

We combine these findings in~\fig{fig:allmolecule:o2}
where we show the stability regions of the all-molecule
state.
We identify three regions (I, II and III)
separated by two boundaries,
$f_{TW}$ between I and II, and
$f_{GS}^2$ between II and III.
We conclude that the all-molecule state of the
ground state is stable only in Region III,
while the twin state does not have stable all-molecule configurations.


\subsection{Summary of AMBEC eigenstates}

\begin{figure}[htp]
  \centering
  \includegraphics[type=pdf,ext=.pdf,read=.pdf,scale=0.5,clip]{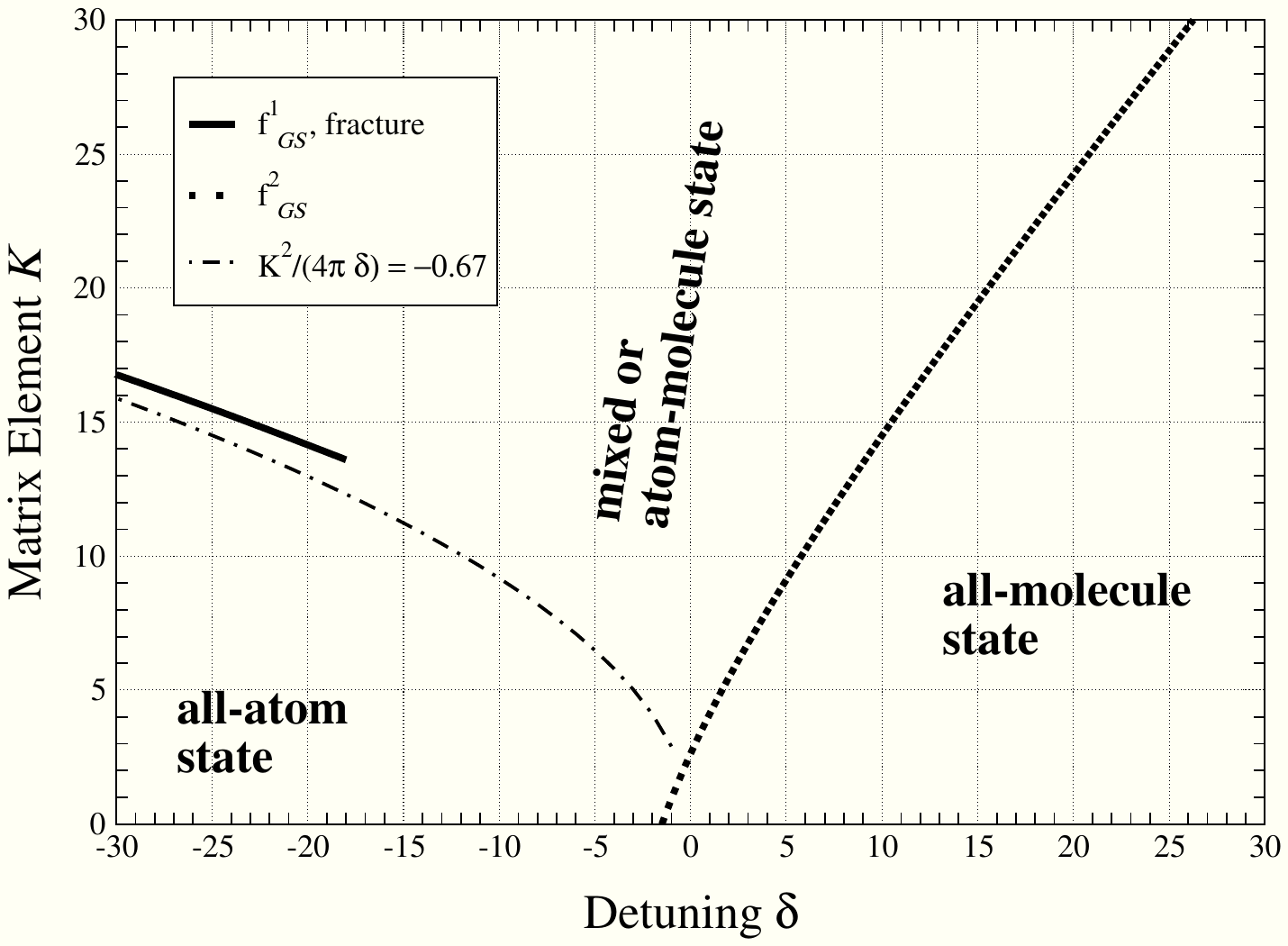}
  \caption{%
    Map of the ground state with the boundaries $f^1_{GS}$ and
    $f^2_{GS}$ separating the three local stationary states.
    Please note that $f^1_{GS}$ only partially bounds the
    all atom state.
    \label{fig:groundstate}
  }
\end{figure}

\begin{figure}[htp]
  \includegraphics[type=pdf,ext=.pdf,read=.pdf,scale=0.5,clip]{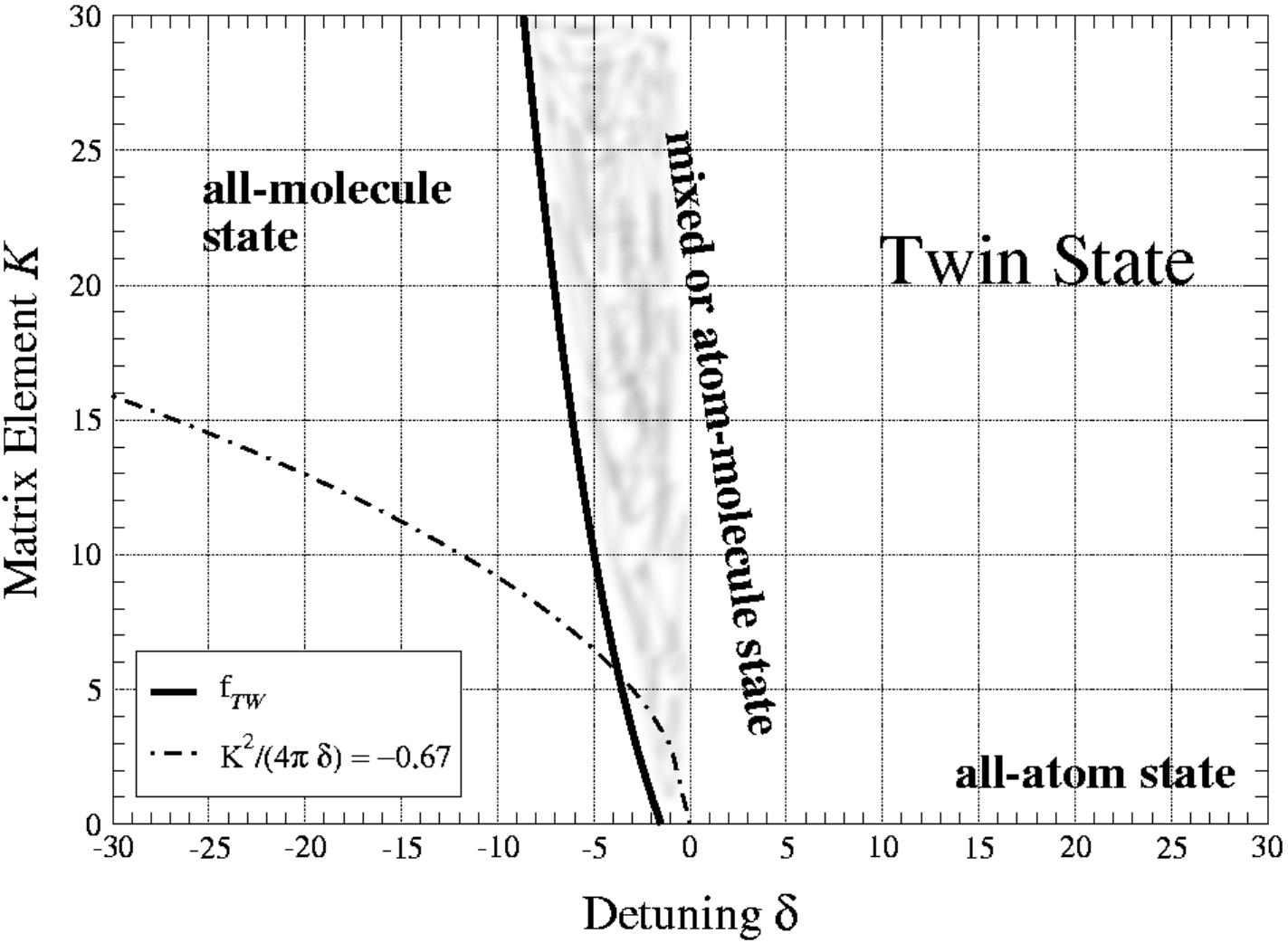}
  \caption{%
    Map of the twin state with the boundary $f_{TW}$
    separating the all-molecule state from the mixed state.
    The actual boundary of the twin
    state is somewhere in the shaded region.
    \label{fig:twinstate}
}
\end{figure}

We construct the maps of the ground state and the twin state from the
analysis of their two limiting configurations, all-atom and all-molecule,
as follows:
\begin{itemize}
  \item {\em Ground state}:
  In \fig{fig:groundstate} we show the outline of the ground state
  proposed in this report.
  It comprises three stationary states:
  all-atom, mixed and all-molecule state.
  The three stationary states are separated by two boundaries,
  $f^2_{GS}$ (between the mixed state and the all-molecule state)
  and $f^1_{GS}$ (partially separates all-atoms from mixed).
  For large negative detunings $f^1_{GS}$ is given by
  the single GPE collapse boundary $\frac{K^2}{4\,\pi\,\delta}=-0.67$.
  \\
  The \ambec\ interprets the collapse as that by crossing $f^1_{GS}$
  for large $K$ by increasing $\delta$,
  the size of atomic mean-field reduces drastically and discontinuously.
  This boundary, however, does not extend to the origin.
  So, at least in principle, it is possible to perform a parameter sweep
  around the boundary and produce controllably narrow atomic mean-field
  without parametric excitations.
\item{\em Twin State}:
  We show its outline in \fig{fig:twinstate}.
  It comprises the all-atom state and the mixed state, and disappears in
  a region of parameter space between $\delta \simeq0$ and $f_{TW}$.
\end{itemize}

\section{Discussion}

We have studied the stationary states of the simple atom-molecule
Bose-Einstein Condensate (AMBEC) numerically and analytically.
We have shown that, while the ground state remains a global eigenstate,
the twin state does not.
Non-linearity of AMBEC equations implies that all stationary states
need not be eigenstates: For that they also have to be stable
under small perturbations.
The boundaries of the stationary states are a likely source of
parametric excitations in sweeps as across such a boundary the properties
of the stationary state may change discontinuously.

In parametric sweeps we have seen two types of parametric excitations
near the all-molecule stationary state.
In the first one type, the fraction of the atomic condensate remains
small, while the size varies erratically.
We saw that in the ground state near the boundary $f_{GS}^2$.
In the second type, the atomic condensate grows in fraction and in
size erratically which in some regions of parameter space may
resemble an explosion.
This is what we see in the decay of the twin state near $f_{TW}$.

A parameter sweep that follows the twin state in parameter space,
has been realized in the ``Bosenova'' experiments~\cite{Donleyetal2001:bosenova}.
The excitations of the BEC in a near-resonant region, observed on that
occasion, followed by the loss of the atomic condensate makes a tempting
parallel to the chaotic behavior we are predicting.
The AMBEC we have examined suggests that the decay occurs
because we have created a system in a local eigenstate, and
have swept the parameters out of the bounds in which the eigenstate
exists.
While the onset of parametric excitations is in the scope of
the mean-field approach,
analysis of its final products may require inclusion of quantum
corrections~\cite{Mackieetal2002:bosenova,Kokkelmansetal2002:bosenova,DuineStoof2003:bosenova,Koehleretal2003:bosenova}.

We have presented a way to calculate system energy even near resonance,
where the standard expansions in $\delta$ fail.
In addition, our result in the case of simple AMBEC hints an exciting possibility
to create almost-all atom state of various sizes while avoiding the parametric excitation tied to
$a_{crit}$ of the all-atom state (as accessible in Feshbach resonance experiments).
Interestingly, our findings suggest that the disintegration of the simple AMBEC twin state
might be even more spectacular.
Studies of a trapped AMBEC on the time scale of the trap, both experimental
and theoretical, might have other such intriguing surprises in store.

\acknowledgements

This work is supported in part by NSF (PHY-0097974) and NASA (NAG8-1428).
\bibliography{%
bibliography/bec,%
bibliography/numericks,%
bibliography/generalphysics%
}

\clearpage
\section*{Figures and Captions}

\begin{figure}[tp]
  \centering
  \includegraphics[type=pdf,ext=.pdf,read=.pdf,scale=1,clip]{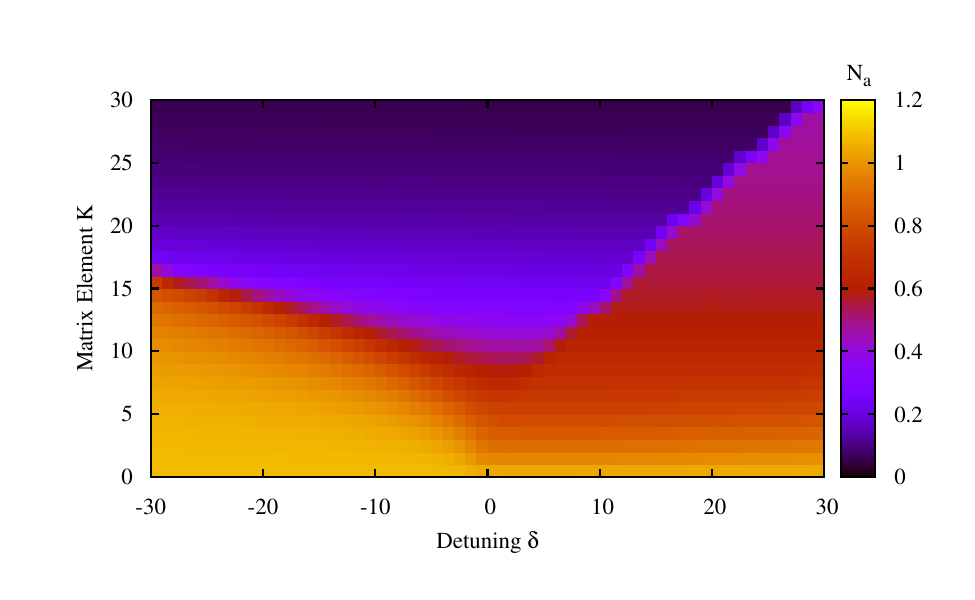}
  \caption{%
    \label{fig:gs:cra}
    (Ground state) $R_{1/2}$ as a function of $\delta$ and $K$ we find
    from evolution of GPEs~\noeq{eq:gpe} in complex time.
    In the region of the mostly-atom state (cf. to \fig{fig:gs:na},in yellow, for $\delta < 0$)
    we see the effects of the boundary $f^1_{GS}$, where it separates the mostly-atom state of
    $R_{1/2} \simeq 1$ from a mixed state which atomic fraction is very narrow, $R_{1/2} \ll 1$.
}
\end{figure}

\clearpage

\clearpage

\clearpage

\clearpage

\clearpage


\clearpage
\section*{Tables and Captions}


\begin{table}[tbph]
  \begin{center}
    \begin{tabular}[c]{|c||c||c|cc|}
      \hline
      $\delta$ & $K_{var}$ & $K_{num}$ & $N_a^{mix}$ & $R_{1/2}^{mix}$\\
      \hline\hline
      -1.5&  0   &  0             &   -  &   -\\
      0   &  2.6 &  2.5 $\pm$ 0.5 & 0.13 & 0.76\\
      1   &  4.1 &  4.5 $\pm$ 0.5 & 0.18 & 0.76\\
      5   &  9.1 &  9.5 $\pm$ 0.5 & 0.25 & 0.52\\
      10  & 14.5 & 14.5 $\pm$ 0.5 & 0.42 & 0.28\\
      20  & 24.2 & 24.5 $\pm$ 0.5 & 0.55 & 0.09\\
      30  & 33.4 & 33.5 $\pm$ 0.5 & 0.66 & 0.03\\
      \hline
    \end{tabular}
    \caption{%
      Comparison of the variational and (full AMBEC) numerical results for the position
      of the boundary $f^2_{GS}$ in the format $K$ for various $\delta$.
      From numerical solutions we also extract the fraction of atoms $N_A$ and the half-radius $R_{1/2}$
      for the mixed state (superscript $mix$) at $K_{num}+0.5$, on the all-atom side these are both unity.
}
    \label{tab:f2gstheoryvsnumericks}
  \end{center}
\end{table}

\begin{table}[ht]
  \begin{center}
    \begin{tabular}[c]{|cc|cc|}
      \hline
      $\delta_{var}$ & $K$ & $ \delta_{num} $ & $ N_a^{mix}$ \\
      \hline\hline
      -1.5  &   0 & - & -\\
      -3.54 &   5 & - & -\\
      -4.98 &  10 & -3.0 & 0.21 \\
      -7.07 &  20 & -3.9 & 0.32 \\
      -8.66 &  30 & - & -\\
      -11.11 &  50 & - & -\\
      -15.42 & 100 & - & -\\
      \hline
    \end{tabular}
    \caption{%
      Comparison of the variational and (full AMBEC) numerical results for the position
      of the boundary $f_{TW}$.
      In column $N_a^{mix}$ we give the estimated fraction of atoms prior to disintegration
      of the condensates.
    }
    \label{tab:ftwtheory}
  \end{center}
\end{table}

\end{document}